\documentclass[%
preprint,
 amsmath,amssymb,
 aps,
prb,
]{revtex4-1}

\usepackage{color}
\usepackage{graphicx}% Include figure files
\usepackage{dcolumn}% Align table columns on decimal point
\usepackage{bm}% bold math
\usepackage{hyperref}% add hypertext capabilities
\usepackage[mathlines]{lineno}% Enable numbering of text and display math
\begin{document}

\preprint{APS/123-QED}

\title{Strong Rashba effect in the localized impurity states of halogen-doped monolayer PtSe$_2$}

\author{Moh. Adhib Ulil Absor}
\email{adib@ugm.ac.id} 
\affiliation{Department of Physics, Universitas Gadjah Mada BLS 21 Yogyakarta Indonesia.}%
%Lines break automatically or can be forced with \\

\author{Iman Santoso}
\affiliation{Department of Physics, Universitas Gadjah Mada BLS 21 Yogyakarta Indonesia.}%

\author{Harsojo}
\affiliation{Department of Physics, Universitas Gadjah Mada BLS 21 Yogyakarta Indonesia.}%

\author{Kamsul Abraha}
\affiliation{Department of Physics, Universitas Gadjah Mada BLS 21 Yogyakarta Indonesia.}%

\author{Hiroki Kotaka}
\affiliation{Elements Strategy Initiative for Catalysts and Batteries (ESICB), Kyoto University, Kyoto 615-8520, Japan
}%

\author{Fumiyuki Ishii}%
\affiliation{Faculty of Mathematics and Physics Institute of Science and Engineering Kanazawa University 920-1192 Kanazawa Japan.}%

\author{Mineo Saito}
\affiliation{Faculty of Mathematics and Physics Institute of Science and Engineering Kanazawa University 920-1192 Kanazawa Japan.}%

\date{\today}% It is always \today, today,
             %  but any date may be explicitly specified

\begin{abstract}
The recent epitaxial growth of 1T-phase of PtSe$_2$ monolayer (ML) has opened a possibility for its novel applications, in particular for spintronics device. However, in contrast to 2H-phase of transition-metal dichalcogenides (TMDs), the absence of spin splitting in the PtSe$_2$ ML may limit the functionality for spintronics application. Through fully-relativistic density-functional theory calculations, we show that large spin splitting can be induced in the PtSe$_2$ ML by introducing a substitutional halogen impurity. Depending on the atomic number ($Z$) of the halogen dopants, we observe an enhancement of the spin splitting in the localized impurity states (LIS), which is due to the increased contribution of the $p-d$ orbitals coupling. More importantly, we identify very large Rashba splitting in the LIS near Fermi level around the $\Gamma$ point characterized by hexagonal warping of the Fermi surface. We show that the Rashba splitting can be controlled by adjusting the doping concentration. Therefore, this work paves a possible way to induce the significant Rashba splitting in the two-dimensional TMDs, which is useful for spintronic devices operating at room temperature.    
\end{abstract}

\pacs{Valid PACS appear here}% PACS, the Physics and Astronomy
                             % Classification Scheme.
\keywords{Suggested keywords}%Use showkeys class option if keyword
                              %display desired											
\maketitle

\section{INTRODUCTION}

Recently, spin-orbit coupling (SOC) plays an important role in widely studied systems including topological insulator \cite {Hasan}, skyrmions \cite {Psaroudaki}, and Rashba materials \cite {Manchon}. When the SOC presences in the crystalline systems with lack of inversion symmetry, an effective magnetic field is generated \cite {Rashba,Dresselhauss}, leading to various physical effects such as current-induced spin polarization \cite {Kuhlen}, the spin Hall effect \cite {Qi}, the spin galvanic effect\cite {Ganichev}, and spin ballistic transport \cite {Lu}, and thus giving rise to practical spintronics device. Especially the Rashba effect \cite {Rashba} attracts considerable attention owing to its electric tunability \cite {Nitta} in a spin field-effect transistor (SFET) \cite {Datta}, as recently realized experimentally \cite {Chuang}. However, for spintronics application,  materials with strong Rashba SOC are highly desirable since they enable us to allow spintronics device operation at room temperature \cite {Yaji}.  

From this perspective, a new class of materials crystallizing in the two dimensional (2D) structures such as transition metal dichalcogenides (TMDs) monolayer (ML) is particularly appealing due to the strong SOC \cite {Zhu,Latzke,Absor4,Absor3}.  The TMDs ML crystallize in a hexagonal structure with $MX_{2}$ stoichiometry, where $M$ and $X$ are transition metal and chalcogen atoms, respectively. Depending on the chalcogen stacking, two different stable formations of the $MX_{2}$ is achieved in the ground state, namely a $H$ phase having a trigonal prismatic hole for metal atoms, and a $T$ phase that consists of staggered chalcogen layers forming an octahedral hole for metal atoms \cite {Cudazzo}. For spintronics application, the $H-MX_{2}$ ML systems such as molybdenum and tungsten dichalcogenides (MoS$_{2}$, MoSe$_{2}$, WS$_{2}$, and WSe$_{2}$) have been widely studied \cite {Zhu,Latzke,Absor4,Absor3}. In these systems, the lack of crystal inversion symmetry together with strong SOC in the 5$d$ orbitals of transition metal atoms leads to spin-valley coupling, which is responsible for the appearance of valley-contrasting effects such as valley-selective optical excitations \cite {LaMountain}, valley Hall response \cite {Cazalilla}, spin-dependent selection rule for optical transitions \cite{YWu}, and magneto-electric effect \cite {Gong}.  

Recently, PtSe$_{2}$ ML,  a new member of the 2D TMDs ML with $T-MX_{2}$ ML structures, has been epitaxially grown successfully on the Pt(111) substrate \cite {LWang}. This material exhibits the largest electron mobility among the studied TMDs ML \cite {XZhang}. In contrast to the $H-MX_{2}$ ML systems, the crystal structure of the PtSe$_{2}$ ML is globally centrosymmetric, but, it has strong local dipole field in the two sub-layers\cite {Yao}. Consequently, the SOC induces local Rashba effect exhibiting hidden spin polarizations\cite {dZhang,Razzoli}, i.e., the spin-polarized states are degenerated in energy but spatially locked into two sub-layers forming an inversion partner, as recently observed experimentally by Yao et al. using spin- and angle-resolved photoemission spectroscopy (spin-ARPES) \cite {Yao}. The observed spin-polarized states without the characteristic of the spin splitting in the PtSe$_{2}$ ML may provide a disadvantage for spintronics device. Therefore, finding a possible way to induces spin splitting in the PtSe$_{2}$ is crucially important, which is expected to enhance the functionality for spintronics application.   
 
In this paper, by using fully-relativistic density-functional theory (DFT) calculations taking into account the effect of the SOC, we show that large spin splitting can be induced in the PtSe$_2$ ML by introducing a substitutional halogen impurity. We find that depending on the atomic number ($Z$) of the halogen dopants, enhancement of the spin splitting is achieved in the localized impurity states (LIS), which is due to the increased contribution of the $p-d$ orbitals coupling. More importantly, we identify very large Rashba splitting in the LIS near Fermi level around the $\Gamma$ point characterized by hexagonal warping of the Fermi surface. We show that this Rashba splitting can be controlled by adjusting the doping concentration. Finally, a possible application of the present system for spintronics will be discussed.

\section{Model and Computational Details}

To investigate the effect of an impurity on the electronic properties of the PtSe$_{2}$ ML, we performed first-principles electronic structure calculations based on the density functional theory (DFT) within the generalized gradient approximation (GGA) \cite {Perdew} using the OpenMX code \cite{Openmx}. We used norm-conserving pseudopotentials \cite {Troullier}, and the wave functions are expanded by the linear combination of multiple pseudoatomic orbitals (LCPAOs) generated using a confinement scheme \cite{Ozaki,Ozakikino}. In the case of the pristine PtSe$_{2}$ ML, the atomic orbitals are specified by Pt7.0-$s^{2}p^{2}d^{2}$ and Se9.0-$s^{2}p^{2}d^{1}$, which means that the cutoff radii are 7.0 and 9.0 Bohr for the Pt and Se atoms, respectively, in the confinement scheme \cite{Ozaki,Ozakikino}. For the Pt atom, two primitive orbitals expand the $s$, $p$, and $d$ orbitals, while, for the Se atom, two primitive orbitals expand the $s$ and $p$ orbitals, and one primitive orbital expands $d$ orbital. The impurity is taken from the halogen family such as F, Cl, Br, and I atoms. Similar to the Se atom, two primitive orbitals expand the $s$ and $p$ orbitals, and one primitive orbital expands $d$ orbital for the halogen atoms. The effect of the SOC was included in our DFT calculations.

\begin{figure}
	\centering
		\includegraphics[width=0.8\textwidth]{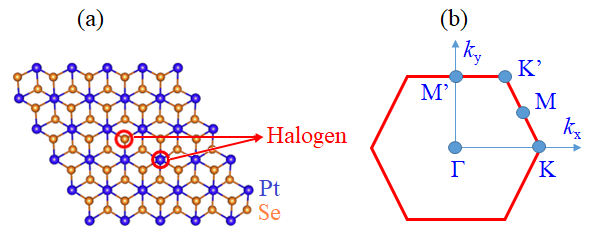}
	\caption{(a) Crystal structures halogen doped PtSe$_2$ ML from top view. The halogen impurity is substituted on the Se site as indicated by red circle. (b) First Brillouin zone of the monolayer indicated by the high symmetry points (K, $\Gamma$, M). Here, the $x$ axis is chosen along the $\Gamma$-K direction. }
	\label{figure:Figure1}
\end{figure}

To model the PtSe$_2$ ML, we used a periodic slab model with a sufficiently large vacuum layer (20 \AA). Next, we construct a 4x4x1 supercell of the PtSe$_{2}$ ML to model the impurity system. We then introduce a substitutional halogen impurity in the PtSe$_{2}$ ML where two different cases are considered: (i) the impurity is located on the Se site and (ii) Pt site [Fig. 1(a)]. The larger supercells (5x5x1 and 6x6x1 supercells) are used to test our calculational results, and we confirmed that it does not affect the main conclusion. The geometries were fully relaxed until the force acting on each atom was less than 1 meV/\AA. To confirm the stability of the impurity, we calculated formation energy of a particular substitutional dopant $E_{f}$ defined as:
\begin{equation}
\label{1}
E_{f}=E_{{PtSe_{2}}:X}-E_{PtSe_{2}}-\mu_{X}+\mu_{Pt(Se)}.
\end{equation}
where $E_{{PtSe_{2}}:X}$ is the total energy of the halogen doped PtSe$_{2}$ ML, $E_{PtSe_{2}}$ is the total energy of the pristin PtSe$_{2}$ ML, while $\mu_{X}$ and $\mu_{Pt(Se}$ are the chemical potential of the substitutional halogen atom and the substituted Se (Pt) host atoms, respectively. Here, both $\mu_{Pt}$ and $\mu_{Se}$ obtain the following requirements:
\begin{equation}
\label{2}
E_{PtSe_{2}}-2E_{Se}\leq \mu_{Pt}\leq E_{Pt},
\end{equation}
\begin{equation}
\label{3}
\frac{1}{2}(E_{PtSe_{2}}-E_{Pt})\leq \mu_{Se}\leq E_{Se}.
\end{equation}
Under Se-rich condition, $\mu_{Se}$ is the energy of the Se atom in the bulk phase (hexagonal Se, $\mu_{Se}=\frac{1}{3}E_{Se-hex}$) which corresponds to the lower limit on Pt, $\mu_{Pt}=E_{PtSe_{2}}-2E_{Se}$. On the other hand, in the case of the Pt-rich condition, $\mu_{Pt}$ is associated with the energy of the Pt atom in the bulk phase (fcc Pt, $\mu_{Pt}=\frac{1}{4}E_{Pt-fcc}$) corresponding to the lower limit on Se, $\mu_{Pt}=\frac{1}{2}(E_{PtSe_{2}}-E_{Pt})$.

\section{RESULT AND DISCUSSION}

Before we discuss the effect of a halogen impurity on the electronic properties of the PtSe$_{2}$ ML, we examine structural and energetic stability. The PtSe$_{2}$ ML belongs to a $T$ structure ($T-MX_{2}$) with $P\overline{3}mI$ space group. However, it has a polar group $C_{3v}$ and a centrosymmetric group $D_{3d}$ for the Se and Pt sites, respectively. Here, one transition metal atom (or chalcogen atom) is located on top of another transition metal atom (or chalcogen atom) forming octahedral coordination, while it shows trigonal structure when projected to the (001) plane [Fig. 1(a)]. We find that the calculated lattice constant of the PtSe$_{2}$ ML is 3.75 \AA, which is in good agreement with the experiment (3.73 \AA \cite {LWang}) and previous theoretical calculations (3.75 \AA \cite {LZhuang,Zulfiqar}).  

\begin{table}[h!]
\caption{Physical parameters of pure and doped monolayer calculated using 4x4x1 supercell. $d_{X-\texttt{Pt}}$ (in \AA) is the bond length between $X$ (Se or halogen) and Pt atoms of the pure or halogen-doped systems. $E_{f}$ (in eV) is the formation energy of halogen substitutional dopants under the Se-rich and Pt-rich conditions.} % title of Table
\centering % used for centering table
\begin{tabular}{c c c c c} % centered columns (4 columns)
\hline\hline %inserts double horizontal lines
  Model &     & Bond lenth   & $E_{f}$ (Se-rich) & $E_{f}$ (Pt-rich) \\ % inserts table %heading
\hline % inserts single horizontal line
Pure     & $d_{\texttt{Se-Pt}}$ & 2.548 &         &        \\ % inserting body of the table
Doping in Se sites    &  &  &         &        \\
F doping  & $d_{\texttt{F-Pt}}$  & 2.402 &  -5.87  & -5.52  \\         
Cl doping & $d_{\texttt{Cl-Pt}}$ & 2.615 & -4.11   & -3.76   \\         				
Br doping & $d_{\texttt{Br-Pt}}$ & 2.715 & -3.11   & -2.77\\				 
I doping  & $d_{\texttt{I-Pt}}$  & 2.843 & -1.13 & -0.78 \\         % [1ex] adds vertical space
Doping in Pt sites    &  &  &         &        \\
F doping  & $d_{\texttt{F-Se}}$  & 2.652 & 0.19    & 0.56  \\         
Cl doping & $d_{\texttt{Cl-Se}}$ & 2.681 & 0.05   &  0.42  \\         				
Br doping & $d_{\texttt{Br-Se}}$ & 2.797 & -0.07    & 0.30\\				 
I doping  & $d_{\texttt{I-Se}}$  & 2.893 & -0.33   & 0.04 \\         % [1ex] adds vertical space
\hline\hline %inserts single line
\end{tabular}
\label{table:Table 2} % is used to refer this table in the text
\end{table}

When a halogen impurity is introduced, the position of the atoms around the impurity site is substantially relaxed from the pristine atomic positions. To examine the optimized structure of the impurity systems, we show the calculated results of halogen-Pt and halogen-Se bond lengths ($d_{\texttt{Hal-Pt}}$,$d_{\texttt{Hal-Se}}$) in Table I. In the case of the impurity on the Se site, three Pt atoms surrounding the impurity site are found to be relaxed. Consequently, the bond length $d_{\texttt{Hal-Pt}}$ becomes smaller or larger than $d_{\texttt{Se-Pt}}$ in the pristine system depending on the halogen atoms. For instant, in the case of F doping, $d_{\texttt{F-Pt}}$ (2.402 \AA) is smaller than $d_{\texttt{Se-Pt}}$ (2.548 \AA) in the pristine system. However, for the case of Cl, Br, and I dopings, $d_{\texttt{Hal-Pt}}$ (2.615 \AA, 2.715 \AA, and 2.843 \AA, respectively) are larger than $d_{\texttt{Se-Pt}}$ in the pristine system. Since $d_{\texttt{Hal-Pt}}$ at each hexagonal side has the same value, trigonal symmetry suppresses the impurity to exhibit the $C_{3v}$ point group, which is similar to those observed on the halogen-doped WS$_{2}$ \cite {Guo} ML and Se vacancy of PtSe$_{2}$ ML \cite {Absor1}. Similarly, the impurity on the Pt site also induces atomic relaxation so that the six Se atoms are significantly moved away from each other. Therefore, $d_{\texttt{Hal-Se}}$ is larger than $d_{\texttt{Se-Pt}}$ in the pristine system [see Table 1]. However, three-fold rotation preserves around the impurity site. Thus the symmetry of the system retains the $D_{3d}$ point group.

\begin{figure*}
	\centering
		\includegraphics[width=1.0\textwidth]{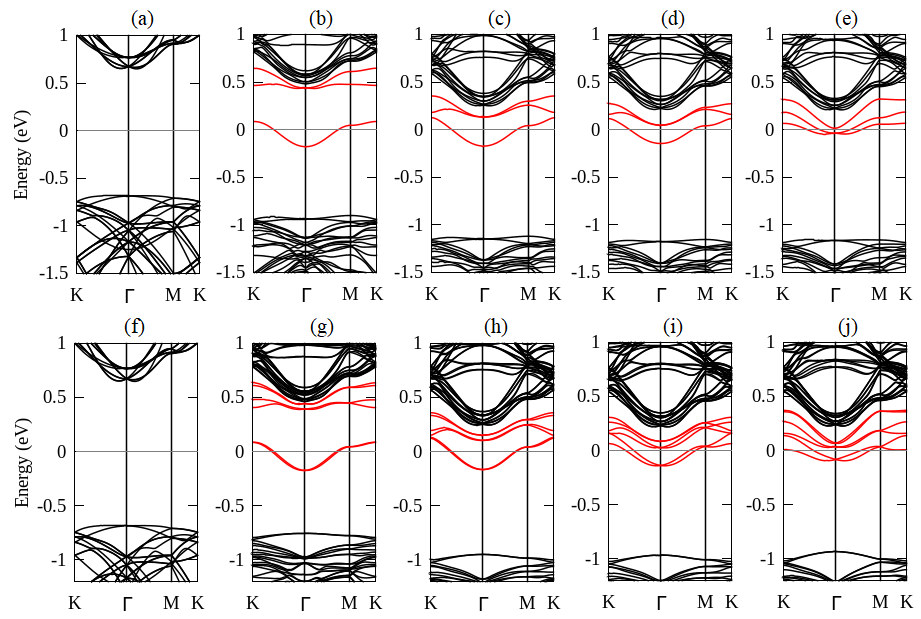}
	\caption{The electronic band structure of (a) the pristine, (b) F-doped, (c) Cl-doped, (d) Br-doped, and (e) I-doped PtSe$_{2}$ML where the calculations are performed without inclusion the effect of the spin-orbit coupling (SOC). The electronic band structure of (f) the pristine, (g) F-doped, (h) Cl-doped, (i) Br-doped, and (j) I-doped PtSe$_{2}$ML with inclusion the effect of the SOC. The Fermi level is indicated by the dashed black lines.}
	\label{figure:Figure2}
\end{figure*}

The significant structural changes induced by a halogen impurity is expected to strongly affect the stability of the PtSe$_{2}$ ML, which is confirmed by the calculated result of the formation energy ($E_{f}$) as given in Table I. We find that $E_{f}$ of the impurity on the Se site is much lower than that on the Pt site, indicating that the formation of the impurity on the Se site is more favorable. Moreover, the calculated value of $E_{f}$ under the Se-rich condition is smaller than that under the Pt-rich condition, showing that the doped compounds under Se-rich can be realized in the equilibrium condition. Furthermore, for the larger $Z$ element of the halogen atoms, the formation of the impurity is stabilized by enlarging the bond length $d_{\texttt{Hal-Pt}}$. Therefore, $E_{f}$ increases from F to I doping systems, which is consistent with that observed on MoS$_{2}$ \cite {Dolui} and WS$_{2}$ \cite {Guo} MLs.       

Strong modification of electronic properties of the PtSe$_{2}$ ML is expected to be achieved by introducing a halogen impurity. Here, we focused on the impurity on the Se site since it has lower formation energy than that on the Pt site. Figure 2 shows electronic band structures of the impurity systems compared with those of the pristine one. In contrast to the pristine system [Figs. 2(a) and 2(f)], we identify localized impurity states (LIS) in the band structures of the impurity systems, which are located close to the conduction band minimum (CBM) [Figs. 2(b)-2(e)]. More importantly, we find spin-split bands at the LIS [Figs. 2(g)-2(j)] when the SOC is taken into account. Depending on the $Z$ number of the halogen dopants, enhancement of the spin splitting in the localized impurity states (LIS) is observed for the larger $Z$ element, indicating that these systems are promising for spintronics application.  

\begin{figure}
	\centering
		\includegraphics[width=1.0\textwidth]{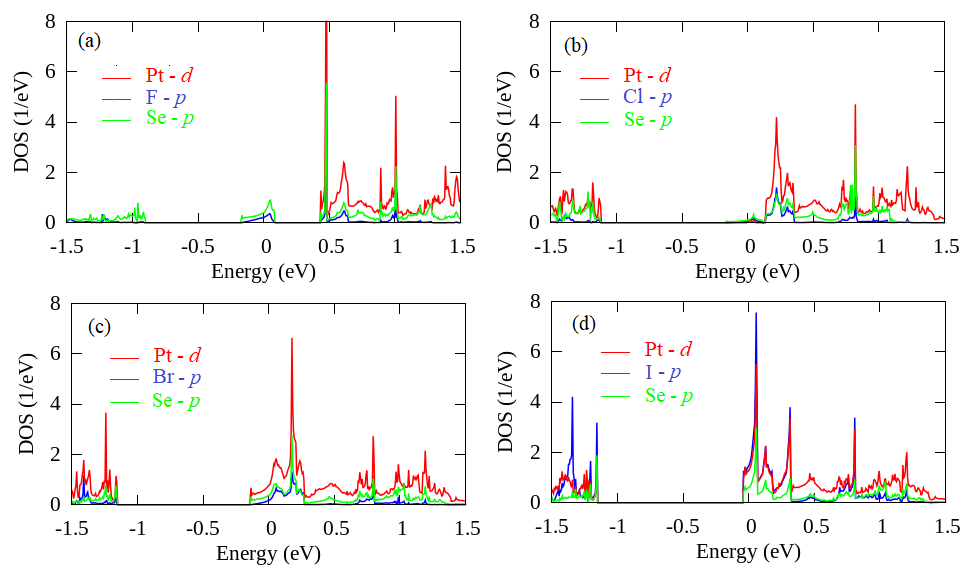}
	\caption{Density of states (DOS) projected to the atomic orbitals calculated around the impurity site for: (a) F-doped, (b) Cl-doped, (c) Br-doped, and (d) I-doped PtSe$_{2}$ ML.}
	\label{figure:Figure3}
\end{figure}

To clarify the origin of the spin-split bands in the LIS, we show in Fig. 3 the calculated results of the density of states (DOS) projected to the atomic orbitals. In the atomic representation, coupling between atomic orbitals will contributes to the non-zero SOC matrix element through relation $\zeta_{l}\left\langle \vec{L}\cdot \vec{S}\right\rangle_{u,v}$, where $\zeta_{l}$ is angular momentum resolved atomic SOC strength with $l=(s,p,d)$, $\vec{L}$ and $\vec{S}$ are the orbital angular momentum and Pauli spin operators, and $(u,v)$ is the atomic orbitals. Accordingly, the orbitals hybridization play an important role in inducing the spin splitting. From the calculated results of the DOS, it is found that different orbitals hybridization in the LIS is observed under different impurity systems. In the case of the F and Cl dopings, we find that the LIS is dominated by coupling between the halogen-$p$ and the nearest-neighbor Se-$p$ orbitals, while the contribution of the Pt-$d$ orbital is small [Figs. 3(a) and 3(b)]. As a result, the SOC matrix element contributes only minimally to the spin splitting [Figs. 2(g) and 2(h)]. However, strong coupling between Pt-$d$, Se-$p$, and halogen-$p$ orbitals is achieved in the LIS of the Br and I dopings [Figs. 3(c) and 3(d)], which is responsible for inducing the large spin splitting in the LIS as shown in Figs. 2(g)-2(j). It is noted here that in all of the impurity systems, the Pt-$d$ orbital is dominant in the CBM. This is, in fact, consistent with the calculated DOS on the pristine system where the Pt-$d$ orbital plays a significant role in characterizing the CBM as shown in Fig. S1 in the supplemental material\cite {Supplemental}.

To further analyze the spin-split bands in the LIS, we discuss our systems in term of symmetry argument. Here, the LIS can be identified according to the group of the wave vector (GWV) at high-symmetry points in the Brillouin zone (BZ). Similar to the space group of the real space, the GWV belongs to $C_{3v}$ at the $\Gamma$ point. Therefore, the absence of the SOC leads to the fact that the LIS are decomposed into singlet and doublet characterized by $A_{1}$ and $E$ of the single-group irreducible representations (IRs), respectively. However, away from the $\Gamma$ point, lowering symmetry of the GWV is achieved, which is expected to induce splitting bands. Taking the $K$ point as an example, the point group of the GWV becomes $C_{3h}$. Here, $A_{1}$ at the $\Gamma$ point transforms into $A'$ at the $K$ point, while $E$ at the $\Gamma$ point splits into $\left\{E',E'^{*}\right\}$ and $\left\{E",E"^{*}\right\}$ at the $K$ point [See Fig. S2 (a) in the supplementl material \cite {Supplemental}]. Introducing the SOC, the spin-split bands of the LIS are established according to double-group IRs of the GWV. The double group IRs are evaluated by the direct product between single group ($\Gamma_{i}$) and spin representation ($D^{1/2}$) through the relation $\Gamma_{i}\otimes D^{1/2}$. At the $\Gamma$ point, the direct product leads to the fact that $A_{1}$ transforms into $\Gamma_{6}$, while $E$ splits into $\Gamma_{6}$, $\Gamma_{5}$, and $\Gamma_{4}$. On the other hand, at the $K$ point, $A'$ splits into $K_{7}$ and $K_{8}$, while $\left\{E',E'^{*}\right\}$ and $\left\{E",E"^{*}\right\}$ splits into $\left\{K_{10},K_{12},K_{11},K_{9}\right\}$ and $\left\{K_{8},K_{11},K_{12},K_{7}\right\}$, respectively. The classification of the double-group IRs in the energy band at the $\Gamma$ and $K$ points is given in Fig. S2 (b) in the supplemental material \cite {Supplemental}. 

Next, we focused on the spin-split bands of the LIS near Fermi level. Here, we choose the I doping as a representative of halogen impurity systems because of the enhanced spin splitting. The band structures near the Fermi level calculated without and with the SOC are shown in Figs. 4(a) and 4(b), respectively. We find that besides a large valley splitting ($\Delta_{Kv}=135$ meV) at the $K$ point, we also observed the obvious Rashba splitting around the $\Gamma$ point [Fig. 4(b)]. To quantify the strength of the Rashba splitting ($\alpha_{R}$), we show in Fig. 4(c) the highlighted spin-split bands characterized by the Rashba energy ($E_{R}$) and momentum offset ($k_{R}$). Here, $E_{R}$ and $k_{R}$ are important to stabilize spin precession and achieve a phase offset for different spin channels in the spin-field effect transistor device \cite{}. We summarize the calculated results of the parameters $E_{R}$, $k_{R}$, and $\alpha_{R}$ in Table II, and compare these results with a few selected systems from previously reported calculations. It is found that the calculated value of $\alpha_{R}$ in the case of the I doping is 1.7 eV\AA, which is the largest among the halogen impurity systems. Moreover, the $\alpha_{R}$ in the case of the I doping is much larger than that of the conventional semiconductor heterostructures InGaAs/InAlAs \cite{Nitta_b}, the oxide interface LaAlO$_{3}$/SrTiO$_{3}$ \cite{Zhong}, surface Au(111) \cite{LaShell}, and Bi(111)\cite{Koroteev}. Even, this value is comparable with the bulk BiTeBr \cite{Sakano}, BiTeCl \cite{Xiang}, GeTe \cite{DiSante}, and SnTe \cite{Plekhanov}, and newly reported 2D materials including LaOBiS$_{2}$ \cite{LiuQ}, and BiSb \cite{Singh}.

\begin{figure*}
	\centering
		\includegraphics[width=1.0\textwidth]{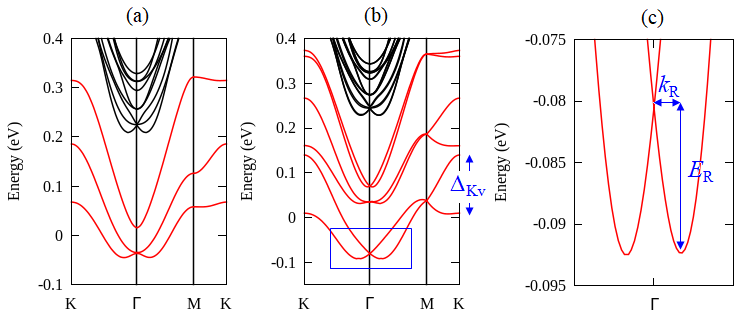}
	\caption{Spin-split bands of the LIS near Fermi level calculated around the $\Gamma$ point: (a) without SOC and (b) with SOC. (c) The highlighted Rashba spin-split bands near degenerated around the $\Gamma$ point. Here, the spin splitting energy at the K poin ($\Delta_{Kv}$), the Rashba energy ($E_{R}$) and momentum offset ($k_{R}$) are indicated. }
	\label{figure:Figure4}
\end{figure*}

\begin{table}[h!]
\caption{Several selected two-dimensional materials and parameters characterizing the Rashba splitting: the Rashba energy $E_{R}$ in meV, momentum offset $k_{R}$ in \AA$^{-1}$, and Rashba parameter $\alpha_{R}$ in eV\AA} % title of Table
\centering % used for centering table
\begin{tabular}{c c c c c} % centered columns (4 columns)
\hline\hline %inserts double horizontal lines
  Systems & $k_{R}$  & $E_{R}$  & $\alpha_{R}$  & Reference \\ % inserts table %heading
\hline % inserts single horizontal line
I doping   & 0.015   & 12.5 &  1.7  & This work  \\        
Br doping  & $9.15 \times 10^{-3}$ & 4.91 & 1.07   & This work   \\         	
Cl doping  & $3.02 \times 10^{-3}$ & $1.17 \times 10^{-5}$ & $5.85 \times 10^{-3}$   & This work\\				 
F doping   & $4.15 \times 10^{-3}$  & $5.18 \times 10^{-5}$ & $10.5 \times 10^{-3}$  & This work \\
Surface  &  &  &   &  \\
Au (111) surface  & 0.012  & 2.1 & 0.33  & Ref.\cite{LaShell} \\
Bi (111) surface  & 0.05  & 14 & 0.55  & Ref.\cite{Koroteev} \\ 
Interface  &  &  &   &  \\
InGaAs/InAlAs  & 0.028  & $<1$ & 0.07  & Ref.\cite{Nitta_b} \\ 
LaAlO$_{3}$/SrTiO$_{3}$  &  & $<5.0$ & 0.01 to 0.05  & Ref.\cite{Zhong} \\ 
Bulk  &  &  &   &  \\
BiTeCl$_{2}$  & 0.03 & 18.45 & 1.2  & Ref.\cite{Xiang} \\
BiTeBr$_{2}$   &  $<0.05$ & $<50$ & $<2$  & Ref.\cite{Sakano} \\
GeTe   &  0.09 & 227 & 4.8  & Ref.\cite{DiSante} \\
SnTe&  0.08 & 272 & 6.8  & Ref.\cite{Plekhanov} \\
2D ML&   &  &   &  \\ 
LaOBiS$_{2}$  & 0.025 & 38 & 3.04  & Ref.\cite{LiuQ} \\ 
BiSb ML  & 0.0113  & 13 & 2.3  & Ref.\cite{Singh} \\          % [1ex] adds vertical space
\hline\hline %inserts single line
\end{tabular}
\label{table:Table 3} % is used to refer this table in the text
\end{table}

It is noted here that the calculated results of $\alpha_{R}$ shown in Table II are obtained from the linear Rashba model where energy band dispersion is written as $E(k)=\frac{\hbar^{2}}{2m^{*}}(\left|k\right|\pm k_{R})^{2}+E_{R}$. Here, $\alpha_{R}$ is expressed as $\alpha_{R}=2E_{R}/k_{R}$, and $m^{*}$ is the electron effective mass. However, for the accuracy of $\alpha_{R}$, we should take into account the higher order correction of $k$ in the Rashba Hamiltonian $H_{\texttt{R}}$. Since the impurity systems have $C_{3v}$ symmetry, the total Hamiltonian $H_{\texttt{T}}$ can be expressed as kinetic part $H_{\texttt{0}}$ and the Rashba part $H_{\texttt{R}}$ up to third order correction of $k$ as \cite{Vajna,Absor1,Absor2} 
\begin{equation}
\label{4}
H_{\texttt{T}}=H_{\texttt{0}}+H_{\texttt{R}}=\frac{\hbar^{2}k^{2}}{2m^{*}}+\left\{(\alpha_{R}k+\beta_{R}k^{3})(\sin\theta \sigma_{y}-\cos\theta\sigma_{x})+\gamma_{R}k^{3} \cos(3\theta)\sigma_{z}\right\},
\end{equation}
where $k=\sqrt{{k^2_{x}}+{k^2_{y}}}$, $\theta =\tan^{-1}(k_{y}/k_{x})$ is the azimuth angle of momentum $k$ with respect to the $x$ axis along the $\Gamma$-K direction, and $\sigma_{i}$ are Pauli matrices. In Eq. (\ref{4}), the parameters $\alpha_{R}$ and $\beta_{R}$ characterize in-plane spin polarizations, while $\gamma_{R}$ is the warping parameters contributing to the out-of-plane component of spin polarizations. Solving the eigenvalues problem involving Hamiltonian of Eq. (\ref{4}), we obtain the spin splitting energy ($\Delta E$) expressed in the square form as follow:
\begin{equation}
\label{5}
(\Delta E)^2=(\alpha_{R}k+\beta_{R}k^{3})^{2}+\gamma_{R}^{2}k^{6}\cos ^{2}(3\theta)).
\end{equation} 
The parameters $\alpha_{R}$, $\beta_{R}$, and $\gamma_{R}$ can be calculated by numerically fitting of Eq. (\ref{5}) to the spin splitting energy along the $\Gamma$-K and $\Gamma$-M directions obtained from our DFT calculations, and we find that $\alpha_{R}$=1.68 eV\AA, $\beta_{R}$=-9.8 eV\AA$^3$, and $\gamma_{R}=18.3$ eV\AA$^3$. We noted here that the calculated values of $\alpha_{R}$ obtained from the higher order correction is fairly agreement with that obtained from the linear Rashba model. However, the large value of $\beta_{R}$ and $\gamma_{R}$ found in the presence system indicates that the contribution of the higher order correction of the $k$ in the $H_{R}$ to the spin-splitting properties of the LIS is significant, in particular for the spin splitting at higher energy level and large wave vector $k$. 

\begin{figure}
	\centering
		\includegraphics[width=0.75\textwidth]{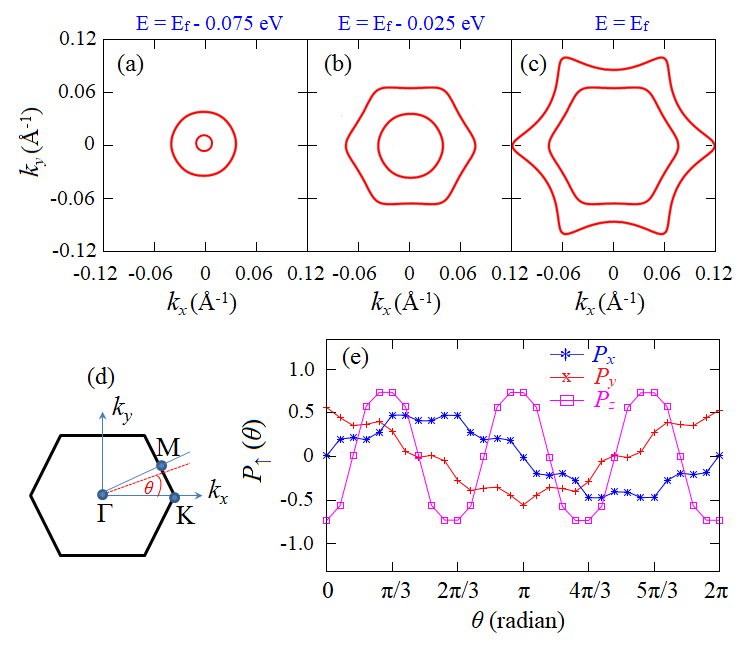}
	\caption{The constant energy contour calculated on : (a) $E=E_{f}-75$ meV, (b) $E=E_{f}-25$ meV, and (c) $E=E_{f}$. (d) Schematic view of spin rotation angle $\theta$ defined in the first Brillouin zone is shown. (e) The calculated result of the spin polarizations for the upper states $P_{\uparrow}$ as a functions of rotation angle $\theta$. Here, $P_{x,y}$ and $P_{z}$ represent the in-plane and out-of-plane spin components, respectively. }
	\label{figure:Figure5}
\end{figure}

To further confirm the significant higher order correction of $k$ to the spin splitting in the LIS, we show a set of a constant energy contour corresponding to the spin-polarized states in the LIS around the $\Gamma$ point near the Fermi level as given in Fig. 5. We find that the shape of constant energy contour has strong energy dependence [Figs. 5(a)-5(c)]. Close to degenerate states around the $\Gamma$ point, we observed circle shape of the energy contour [Fig. 5(a)], but the shape becomes hexagonal at higher energy level [Fig. 5(b)]. Interestingly, we identify the hexagonal warping character of the energy contour at Fermi level [Fig. 5(c)], exhibiting anisotropic Fermi surface. The evolution of the energy contour concerning the energy is similar to those observed on the surface states of Bi$_{2}$Te$_{3}$ surface \cite{LFu}. Moreover, by investigating angle-dependent of the spin polarization ($P_{\uparrow\downarrow}(\theta)$ in the $k$-space [Fig. 5(d)], we identify significant out-of-plane spin components ($P_{z}$) in the Fermi surface along the $\Gamma$-K direction. Here, we find three-fold symmetry of $P_{z}$ with up and down spin alternations [Fig. 5(e)], which is consistent with the $\cos 3\theta$ term in the Eq. (\ref{4}). On the other hand, according to the first term of the $H_{R}$ in the Eq. (\ref{4}), the in-plane spin components ($P_{x}$, $P_{y}$) retains along the $\Gamma$-M direction, inducing helical characters of the spin polarizations [Fig. 5(e)]. Remarkably, the observed out-of-plane spin polarizations together with the helical character of the in-plane spin polarizations in the present system play an important role for controlling the spin precession \cite{Altmann,Kunihashi}, which is significant for generating spin-polarized currents in the spintronics device.

\begin{figure}[h!]
	\centering
		\includegraphics[width=1.0\textwidth]{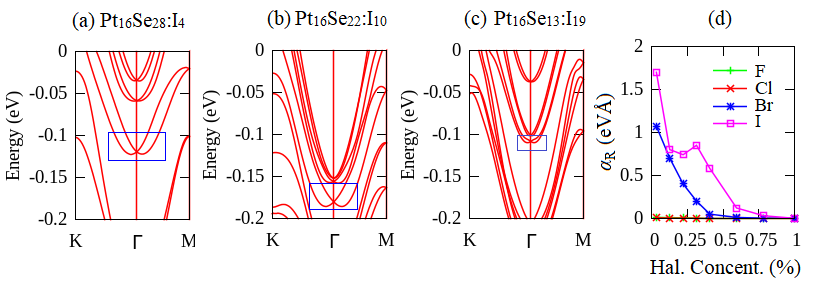}
	\caption{Band structures of the impurity systems with various concentration: (a) Pt$_{16}$Se$_{28}$:I$_{4}$, (b) Pt$_{16}$Se$_{22}$:I$_{10}$, and (c) Pt$_{16}$Se$_{13}$:I$_{19}$. (d) The calculated Rashba spin-orbit strength as a function of dopant concentration. Here, the concentration of the halogen dopant is calculated by using ratio bentween the number of halogen atom per total Se site in the supercell. }
	\label{figure:Figure6}
\end{figure}
 
Next, we discuss the properties of the Rashba splitting in the impurity systems by considering the effect of the doping concentration. Similar to the case of single doping, the Rashba splitting is also visible in the case of multiple dopings, which is observed in the LIS near Fermi level around the $\Gamma$ point [Figs. 6(a)-(6c)]. We also find that the Rashba splitting tends to reduce as the doping concentration enhances, which is, in fact, consistent with decreasing the Rashba spin-orbit strength $\alpha_{R}$ shown in Fig. 6(d). Therefore, the Rashba splitting in the impurity systems can be controlled by adjusting the doping concentration. It is expected that the zero (or very small) Rashba splitting can be achieved when the Se atom in the PtSe$_{2}$ ML is fully replaced by the halogen atoms. Here, a new Pt$X_{2}$ structure, where $X$ is the halogen atom, is achieved, leading to the fact that the symmetry of the system returns to the centrosymmetric group of $D_{3d}$. When the Rashba splitting is extremely small, we can achieve a very long spin coherence. Thus this system can be used as an efficient spintronics device.

Thus far, we have found that the large Rashba spin splitting is achieved on the halogen-doped PtSe$_{2}$ ML. Because the Rashba spin splitting is achieved on the LIS near Fermi level [Fig. 3(c)], $n$-type doping for spintronics is expected to be realized. Therefore, it enables us to allow operation as a spin-field effect transistor device at room temperature \cite {Yaji,Yan,Danker}. We expect that our method for inducing and controlling the large Rashba splitting by using the halogen impurity can also be achieved on other 2D TMDs ML having $T-MX_{2}$ ML systems such as the other platinum dichalcogenides (PtS$_{2}$, PtTe$_{2}$) \cite {Manchanda}, vanadium dichalcogenide (VS$_{2}$, VSe$_{2}$, VTe$_{2}$) \cite {Cudazzo}, stanium dichalcogenide (SnS$_{2}$, SnSe$_{2}$\cite {Gonzalez}, and rhenium dichalcogenides (ReS$_{2}$, ReSe$_{2}$, ReTe$_{2}$) \cite {Horzum} where the structural, symmetry, and electronic properties are similar. Therefore, this work provides a possible way to induce the large Rashba spin splitting in the 2D nanomaterials, which is very promising for future spintronics application.

\section{CONCLUSION}

We have investigated the effect of a substitutional halogen impurity on the electronic properties of the PtSe$_{2}$ ML by employing the first-principles DFT calculations. Taking into account the effect of the SOC, we found that the large spin splitting is observed in the localized impurity states (LIS) near the Fermi level. We also found
that depending on the $Z$ number of the halogen dopants, enhancement of the spin splitting is achieved in the LIS, which is due to the increased contribution of the $p-d$ orbitals coupling. Importantly, we observed very large Rashba splitting in the LIS near Fermi level around the  $\Gamma$ point exhibiting the hexagonal warping character of the Fermi surface. We showed that the Rashba spin-orbit strength could be controlled by adjusting the doping concentration. Recently,
the doped TMDs ML has been extensively studied \cite {Guo,Ma,Noh}. Our study clarified that the halogen doping plays an important role for inducing the strong Rashba effect in the electronic properties of the PtSe$_{2}$ ML, which is useful for designing future spintronics device operating at room temperatures.

\begin{acknowledgments}

This work was partly supported by PDUPT Research Grant (2018) funded by the ministry of research and technology and higher education (RISTEK-DIKTI), Republic of Indonesia. Part of this research was supported by BOPTN Research Grant (2018) founded by Faculty of Mathematics and Natural Sciences, Universitas Gadjah Mada. The computations in this research were performed using the high-performance computing facilities (DSDI) at Universitas Gadjah Mada.
\end{acknowledgments}

\bibliography{Reference}% Produces the bibliography via BibTeX.

%\bibliography

\end{document}